\documentclass[conference]{IEEEtran}
\usepackage[letterpaper, left=0.64in, right=0.64in, top=0.7in, bottom=0.5in, includefoot]{geometry}
\normalsize
\usepackage{graphicx}
\usepackage{subcaption}
\usepackage{hyperref}
\usepackage{url}
\usepackage{array}
\usepackage[utf8]{inputenc} 
\usepackage[T1]{fontenc}
\usepackage{amsmath}
\usepackage{mathtools}
\usepackage{blkarray}
\usepackage{amsfonts}
\usepackage{amssymb}
\usepackage{amsthm}
\usepackage{enumerate}
\usepackage{cite}
\usepackage{lipsum}
\usepackage{mathtools}
\usepackage{cuted}
\usepackage{calc}
\usepackage{color}
\usepackage{epsfig}
\usepackage{setspace}
\usepackage{pstricks}
\usepackage{cancel}
\usepackage{multirow}
\usepackage{mathrsfs}
\usepackage{algorithm}
\usepackage{algpseudocode}
\usepackage{multirow}
\usepackage{diagbox}
\usepackage{multicol}
\usepackage{booktabs,makecell}
\usepackage{mathrsfs}
\usepackage{enumitem}
\newtheorem{defn}{Definition}
\newtheorem{thm}{{\cal T}heorem}

\newtheorem{example}{Example}

\newcommand{\W}{\mathcal{W}}

\newcommand{\A}{\mathcal{A}}

\newcommand{\T}{\mathcal{T}}
\newcommand{\F}{\mathcal{F}}

\newcommand{\U}{\mathcal{U}}
\newcommand{\X}{\mathcal{X}}
\newcommand{\s}{\mathcal{S}}

\input{epsf.sty}

\title{Communication-Efficient  Distributed Computing Through Combinatorial Multi-Access Models}

\begin{document}
\author{
	\IEEEauthorblockN{Shanuja Sasi$^{1}$ and Onur Günlü$^{1,2}$\\}
	\IEEEauthorblockA{
		$^{1}$ Information Theory and Security Laboratory (ITSL),
		Linköping University, Sweden \\ $^{2}$ Lehrstuhl für Nachrichtentechnik, TU Dortmund, Germany\\
		E-mail: $\{$shanuja.sasi, onur.gunlu$\}$@liu.se}
}
\maketitle
\thispagestyle{empty}

\begin{abstract}
 This paper explores the  multi-access distributed computing (MADC) model, a novel distributed computing framework where mapper and reducer nodes are distinct entities. Unlike traditional MapReduce frameworks, MADC leverages coding-theoretic techniques to minimize communication overhead without necessitating file replication across mapper nodes. We introduce a new approach utilizing combinatorial designs, specifically 
 $t$-designs, to construct efficient coding schemes that achieve a computation load of 1. By establishing a connection between $t$-designs and MapReduce Arrays, we  characterize the achievable communication loads and demonstrate the flexibility of our method in selecting the number of reducer nodes. The proposed scheme significantly reduces the number of reducer nodes relative to existing combinatorial topology schemes, at the expense of increased communication cost. 
\end{abstract}
\section{introduction}
\label{intro}
{\it Distributed computing (DC)} frameworks, such as Hadoop MapReduce \cite{Mapreduce} that are extensively utilized by companies like Google, Facebook, and Amazon, decompose computational workloads into multiple parallel tasks and distribute them across a network of nodes. 
The MapReduce framework is a widely used system for handling large-scale data processing tasks. It operates in three primary phases: \textit{Map}, \textit{Shuffle}, and \textit{Reduce}. Initially, input data blocks (or files) are replicated across several computing nodes. During the \textit{Map} phase, each node processes the locally stored data to generate \textit{intermediate values (IVs)}. These IVs are then exchanged among nodes in the subsequent \textit{Shuffle} phase, facilitating the computation of the final output functions in the \textit{Reduce} phase.

Coding-theoretic methods have been widely applied in DC to areas such as distributed storage, caching, coded matrix computations, and gradient calculations. Despite their versatility, many of these techniques are designed for specific applications. The study in \cite{LMA} introduced the concept of \textit{coded distributed computing (CDC)}, which utilizes coding strategies during the Shuffle phase and can be integrated into any MapReduce-based framework.
 By leveraging coding in the Shuffle phase, CDC significantly reduces communication overhead compared to uncoded approaches, with the reduction being proportional to the computation load in the Map phase. To enable coding during the Shuffle phase, CDC replicates data files across nodes in the Map phase.
The study in \cite{YYW} introduces a coded computing scheme based on \textit{placement delivery array (PDA)} designs, that captures the trade-off between storage, computation, and communication. 
CDC has been the focus of extensive research \cite{SFZ,YL,DPYTH,LCW,WCJ,LMAFog,PLSSM,WCJnew}, with many CDC studies leveraging PDA-based designs \cite{JQ,YTC,SG,SG2,SG3,SG4}. PDAs, originally proposed in \cite{YCTCPDA} to address coded caching problems, have since been widely utilized in diverse applications beyond their initial scope.

A novel framework, referred to as \textit{multi-access distributed computing (MADC)}, is introduced in \cite{BP}. Unlike the traditional approach in \cite{LMA}, where mapper and reducer nodes are assumed to be the same, the model in \cite{BP} distinguishes between these two types of nodes. Mapper nodes are responsible for storing input data and generating IVs, while reducer nodes gather IVs from connected mapper nodes, exchange IVs with other reducers, and compute the final output functions. The work in \cite{BP} studies a scenario where reducer nodes are connected to mapper nodes via a \textit{combinatorial topology (CT)}. In this topology, each reducer node is uniquely connected to $\alpha$ mapper nodes, ensuring that each group of $\alpha$ mapper nodes corresponds to exactly one reducer node.
In \cite{SGR}, the authors propose a two-layered bipartite graph, termed the \textit{MapReduce graph (MRG)}, along with a related array structure known as the \textit{MapReduce array (MRA)}, to model MADC frameworks. 
The authors further introduce a specific class of MRGs called \textit{nearest neighbor connect-MRGs (NNC-MRGs)}.

 The main contributions of this paper are  summarized as follows. 
It is observed from \cite{SGR} that a single MRA can correspond to multiple MRGs, all of which share the same number of files and reducer nodes. However, these MRGs differ in terms of the number of mapper nodes, the computation load, and the number of mapper nodes each reducer node is connected to. Despite these differences, the communication load remains consistent across all MRGs.
Since our objective is to minimize the computation load as well, we prioritize the MRG with the lowest computation load, ideally one that avoids file replication across mapper nodes. In the traditional MapReduce framework, this is not feasible, as the core principle of CDC relies on file replication to reduce communication load. However, in the MADC model, this is achievable. By leveraging the connection between mapper and reducer nodes, we can reduce communication load without requiring file replication, which is a key advantage of the MADC model.
Thus, we focus on minimizing communication load, specifically by eliminating file replication. To achieve this, we utilize new topologies using a combinatorial design known as $t$-design \cite{stinson}, which governs the connection between mapper and reducer nodes. We then derive coding schemes for these models. A major advantage of using $t$-designs is that the number of reducer nodes does not need to grow exponentially with the number of mapper nodes as in CT. 
This method not only preserves communication efficiency but also allows exploration of a wider variety of network topologies.

{\it Notation:}  The bit wise exclusive OR (XOR) operation is denoted by $\oplus.$ The notation $[n]$ represents the set $\{1,2, \ldots , n\}$. 
A \( t \)-subset \( \T \) of $[\Lambda]$ is a subset of the set \( [\Lambda] = \{1, 2, \ldots, \Lambda\} \) such that \( |\T| = t\), where \( |\T| \) denotes the number of elements in \( \T \). We represent any set of integers, such as $\{1,2,3\}$,  more simply as 
 $\{123\}$ by omitting the commas between the numbers.

\section{Problem Definition}
\label{problem defintion}
\begin{figure}
	\setlength{\belowcaptionskip}{-16pt} 
		\centering
		\includegraphics[scale=0.4]{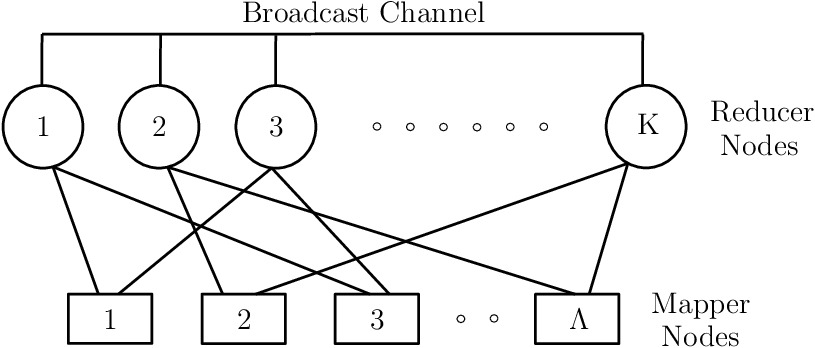}
		\caption{MADC Model.}
		\label{fig: madc model}
\end{figure}
In the MADC model (as shown in Fig. \ref{fig: madc model}) with  a MapReduce framework \cite{BP}, there are $\Lambda$ mapper nodes indexed by $[\Lambda]$, and $K$ reducer nodes indexed by $[K]$. Each reducer node $k \in [K]$ is assigned to compute some output functions which depend on $N$ input files. Thus, the problem is to compute $Q$ output functions, denoted as  $\{\phi_q:  q \in [Q]\}$, from $N$ input files, denoted as $\{w_n : n \in [N]\}$, where the task is distributed across $K$ reducer nodes. Each file $w_n \in {\F}_{2^d}$ with $n \in [N]$ consists of $d$ bits and each function $\phi_q$ for $ q \in [Q]$ maps all  $N$ input files into a stream of $b$ bits, i.e., we have 
$	\phi_q : {\F}_{2^d}^N \rightarrow {\F}_{2^b}.$
We assume that there is a {\it map function} $g_{q,n} : {\F}_{2^d} \rightarrow {\F}_{2^{\beta}}$ for each $n \in [N]$, which maps the input file $w_n$ into an IV $v_{q,n} = g_{q,n}(w_n) \in {\F}_{2^{\beta}}$ of $\beta$ bits, and a {\it reduce function} $h_q : {\F}_{2^{\beta}}^N \rightarrow {\F}_{2^b}$ which maps all IVs into the output value $h_q(v_{q,1}, \ldots , v_{q,N} ) \in {\F}_{2^b}$ of b bits. Thus, $\phi_q$ can be described as 
\begin{align}
	\phi_q(w_1, \dots , w_{N}) = h_q(v_{q,1}, \ldots , v_{q,N} ),   \quad \forall q \in [Q].
\end{align}
Each reducer node $k \in [K]$ is connected to some $\alpha$ mapper nodes and is assigned a subset of the output functions, $\mathcal{W}_k \subseteq [Q]$, where $\mathcal{W}_k$ contains the indices
of the functions assigned to the reducer node $k$. We assume that there is a uniform assignment, which implies $|\mathcal{W}_k| = Q/K $ and $|\mathcal{W}_{k_1} \cap \mathcal{W}_{k_2}| = 0$, for all $k_1,k_2 \in [K]$ such that $k_1$ $ \neq k_2$.
The computation is carried out in three phases:
\begin{enumerate}
	\item {\bf Map Phase:}  The files are divided by grouping  $N$ files into $F$ disjoint batches $\mathcal{B} =\{B_{1},B_{2},\ldots, B_{F}\}$, each containing $\eta_1= N / F$ files such that $\bigcup_{m=1}^{F} B_{m} = \{w_1,w_2,\ldots,w_{N}\}$. Each mapper node $\lambda \in [\Lambda]$ locally stores a subset of batches 
	$M_{\lambda} \subseteq \mathcal{B}$
	and computes the set 
	$	\{v_{q,n} \!=\! g_{q,n}(w_n)\!:\! q \in [Q], w_n \in B_{f},B_{f} \in M_{\lambda}, f \in [F]\}$,
	where each $v_{q,n}$ is a bit stream of length $\beta$ and is referred to as an IV.
	\item {\bf Shuffle Phase:} Each reducer node $k \in [K]$ is connected to some $\alpha$ mapper nodes and can access all files which those mapper nodes have and retrieve the IVs from those mapper nodes. Each reducer node $k$ creates a sequence
		${\bf {X}}_k \in {\F}_{2^{l_k}}$
	and multicasts it to all other reducer nodes via the broadcast link which connects the reducer nodes. We assume that each reduce node receives all the multicast transmissions without any error.
	\item {\bf Reduce Phase:} Recall that each reducer node $k \in [K]$ is assigned a subset of output functions whose indices are in $\mathcal{W}_k$ and requires to recover the IVs 
	$	\{v_{q,n} : q \in \mathcal{W}_k, n \in [N]\}$
	to compute $\phi_q$, for each $q \in \mathcal{W}_k$. Receiving the sequences ${\{{\bf X}_j\}}_{j \in [K] \backslash k}$, each reducer node $k$ decodes all IVs $v_{q,n}$ of its output functions with the help of the IVs it has access to, and finally computes the output functions assigned to them.
\end{enumerate}
\begin{defn}
	(Computation Load \cite{BP}): Computation load $r$ is defined as the total number of files mapped across  $\Lambda$ mapper nodes normalized by the total number of files.
\end{defn}
\begin{defn}
	(Communication Load \cite{BP}): The communication load $L$ is defined as the total number of bits transmitted by  $K$ reducer nodes over the broadcast channel during the Shuffle phase normalized by the number of bits of all IVs.
\end{defn}
\subsection{MapReduce Array}
MapReduce Arrays, introduced in \cite{SGR}, are designed to provide coding schemes for MADC models. Their structure is inspired by the PDAs used in the coded caching literature.
\begin{defn} {\bf (MapReduce Array) \cite{SGR}}
	For positive integers $K, F,$ and $S,$ an $F \times K$ array $P = [p_{f,k}]$ with $ f \in [F],$ and $ k \in [K]$ composed of a specific symbol $*$ and $S$ non-negative integers $[S],$ is called a $(K, F, S)$ MapReduce Array (MRA) if it satisfies the following conditions: 
	\label{def:GPDA}
	\begin{itemize}
		\item {\it C1:} Each integer occurs more than once in the array;
		\item {\it C2:} For any two distinct entries $p_{f_1,k_1}$ and $p_{f_2,k_2}, s=p_{f_1,k_1} = p_{f_2,k_2} $ is an integer only if
		\begin{enumerate}
			\item $f_1$ $\neq f_2$ and $ k_1$ $\neq k_2,$ i.e., they lie in distinct rows and distinct columns; and
			\item $p_{f_1,k_2} = p_{f_2,k_1} = *,$ i.e., the corresponding $2 \times 2$ sub-array formed by rows $f_1, f_2$ and columns $k_1, k_2$ must be either of the following forms
			$ \begin{pmatrix}
				s & *\\
				* & s
			\end{pmatrix} $ or 
			$\begin{pmatrix}
				*& s\\
				s & *
			\end{pmatrix}.$
		\end{enumerate}
	\end{itemize}
\end{defn}
\begin{defn}  {\bf ($g$-regular MRA) \cite{SGR}}
	An array $P$ is said to be a $g$-regular
	$(K, F, S)$ MRA or $g$-$(K, F, S)$ MRA if it satisfies {\it C2} and the following condition
	\begin{itemize}
		\item {\it $C1'$:}  Each integer appears $g$ times in $P$, where $g \geq 2$ is a constant.
	\end{itemize}
	\label{def:g-SRA}
\end{defn}

\subsection{Combinatorial Designs}
\begin{defn}
	{\bf (Design) \cite{stinson}}. A design is a pair $(\mathcal{X}, \mathcal{A})$ such that the following properties are satisfied:
	\begin{enumerate}
		\item $\X$ is a set of elements (called points); and
		\item $\A$ is a collection of non-empty subsets of $\X$ (called blocks).
	\end{enumerate}
\end{defn}
\begin{defn}
	\label{def: t design}
	{\bf ($t$-design) \cite{stinson}}.  A $t$-$(\Lambda, \alpha, m)$ design is a design $(\mathcal{X}, \mathcal{A})$, for some positive integers $\Lambda, \alpha, m,$ and $t$ such that $\Lambda > \alpha \geq t$ and the following properties are satisfied:
	\begin{enumerate}
		\item $|\X|=\Lambda$;
		\item each block contains exactly $\alpha$ points, i.e., we have $|A|=\alpha, \forall A \in \A$; and
		\item every $t$-subset of $\X$ is contained in exactly $m$ blocks.
	\end{enumerate}
\end{defn}
A $t$-$(\Lambda, \alpha, m)$ design without repeated blocks is called a simple $t$-$(\Lambda, \alpha, m)$ design.  The number of blocks in a $t$-$(\Lambda, \alpha, m)$ design is $|\A|=\frac{m{\Lambda \choose t}}{{\alpha \choose t}}$  and 
each point occurs in $m_t=\frac{m{\Lambda -1\choose t-1}}{{\alpha-1 \choose t-1}}$ blocks. These $t$-designs have been widely adopted in coded caching literature \cite{design1,design2}.
\section{Main Results}
In this section, we derive MADC topologies using $t$-designs. Using the properties of $t$-designs, we propose a coding scheme with a computation load of 1. First, we construct a MRA from a $t$-design using {\bf Algorithm \ref{algo1}} given below and establish the connection between $t$-designs and MRAs in {\bf Theorem \ref{thm1}}. Leveraging this connection, we derive the achievable communication load for the MADC model in {\bf Theorem \ref{thm2}}.
The proof of {\bf Theorem \ref{thm1}}, which includes the proof of correctness of {\bf Algorithm \ref{algo1}}, is provided in Section \ref{proof thm1}, and the proof of {\bf Theorem \ref{thm2}} is provided in Section \ref{proof thm2}.
\begin{algorithm}
	\caption{An $ {\Lambda } \times {\Lambda \choose t}$  array $P_{\Lambda,t}$ construction from a $t$-$(\Lambda, \alpha, 1)$ design $([\Lambda], \mathcal{A})$, for  positive integers $\Lambda,t,$ and $\alpha$.}
	\label{algo1}
	\begin{algorithmic}[1]
		\Procedure{{\bf 1}: }{}
		Arrange all subsets of size $t +1$ from $[\Lambda]$ in lexicographic order and for any subset $T'$ of size $t+1$, define $y_{t+1}(T')$ to be its order.
		\EndProcedure \textbf{ 1}
		\Procedure{{\bf 2}: }{}Obtain an array  $P_{\Lambda,t}$ of size ${\Lambda } \times {\Lambda \choose t}$.
		Denote the rows by  $ [\Lambda]$ and columns by  $\mathcal{K}=\{(A,U): A \in \A, U \subseteq A, |U| = t\}$. Define each entry $p_{\lambda,(A,U)}$ corresponding to the row indexed by $\lambda$ and the column indexed by $(A,U)$, as
		\begin{align}
		\label{Dk}
		p_{\lambda,(A,U)} = \left\{
		\begin{array}{cc}
		*, &  \text{if } |\lambda \cap A| \neq 0 \\
		y_{t+1}(\lambda \cup U), &  \text{otherwise}
		\end{array} \right\}.
		\end{align}
		\EndProcedure \textbf{ 2}
	\end{algorithmic}
\end{algorithm}
\begin{thm}
	\label{thm1}
	The $\Lambda \times {\Lambda \choose t}$ array obtained using {\bf Algorithm \ref{algo1}} from a $t$-$(\Lambda, \alpha, 1)$ design $([\Lambda], \mathcal{A})$, for some positive integers $\Lambda,t,$ and $\alpha$, is a $(t+1)$-$ \left (  {\Lambda \choose t},\Lambda,  {\Lambda \choose t} (  \frac{\Lambda-\alpha}{t+1} ) \right )$  MRA.
\end{thm}
\begin{thm}
	\label{thm2}
	Given a $t$-$(\Lambda,\alpha,1)$ design  $([\Lambda],\A)$, for positive integers $\Lambda,\alpha,$ and
	$t$, there exists a coding scheme for a MADC model having $\Lambda$ mapper nodes, indexed by $[\Lambda]$, and $K = |\A|$ reducer nodes, denoted by $\{A \in \A\}$,  with an MRG which consists of
	\begin{enumerate}
		\item $\Lambda$ batches of files, indexed by $\{B_1,B_2,\ldots,B_{\Lambda}\}$;
		\item Each mapper node $\lambda \in [\Lambda]$  assigned a batch $M_{\lambda} = \{B_{\lambda}\}$;
		and
		\item Each reducer node $A \in \A$  connected to mapper nodes in the set
		$ \{\lambda : \lambda \in \A\}.$
	\end{enumerate}
	For the corresponding MRG, the computation load is $r=1$, and the communication load achievable is given by
	\begin{align}
		L(1) = \frac{\Lambda-\alpha}{\Lambda t} .
	\end{align}
\end{thm}
\noindent Now, we illustrate {\bf Theorem \ref{thm2}} using an example.
\begin{example}
	\label{exmp1}
Consider a MADC model with \(N = 7\) input files \(\{w_n: n \in [7]\}\) and \(Q = 7\) output functions \(\{\phi_q: q \in [7]\}\) to be computed.  
We are given a \(2\)-\((7,3,1)\) design \(([7], \A)\), where  
$\A \!=\! \{ \{123\}, \{1 45\}, \{167\}, \{246\},
\{257\}, \{347\}, \{356\} \}.$	Note that, for simplicity, we omit commas between numbers within the elements of each set throughout this paper.
All subsets of size \(3\) from the set \([7]\) are ordered in lexicographic as follows:  
$\{123\},\{124\},\{125\},\{126\},\{127\}, \{134\},
\{135\},$\ and so on.
For each of these subsets, we define a function \(y_3(.)\) as follows:  
$
y_3(\{123\}) \!=\! 1,  y_3(\{124\}) \!=\! 2,  y_3(\{125\}) \!=\! 3,  y_3(\{126\}) \!=\! 4,$  and so on.
We then construct an array of size \(7 \times \binom{7}{2} \) with columns indexed as \(\{(A,U): A \in \A, U \subseteq A, |U| = 2\}\) and rows indexed by \([7]\). The entries of the array are filled according to the following rules (using {\bf Algorithm \ref{algo1}}):
\begin{itemize}
	\item If the row index overlaps with the first set in the ordered pair of column indices (A), we place a \( * \) in the corresponding entry.
	\item Otherwise, we take the union of the row index (a single element) with the second set in the ordered pair of column indices (a 2-subset U), forming a 3-subset. The corresponding value of \(y_3(.)\) for that subset is then placed as the entry.
\end{itemize}
Following this procedure, we obtain the array \(P_{7,2}\), given in (\ref{eg algo 1 pda}). This array is a \(3\)-\((21,7,28)\) MRA.  
The integers missing in the array are \(1,10,15,21,24,28,\) and \(29\), which correspond to $y_3(\{123\}), y_3(\{145\}), y_3(\{167\}), y_3(\{246\}), y_3(\{257\}),$ $y_3(\{347\}),$ and \(y_3(\{356\})\), respectively. This means that the value of \(y_3(A)\) is missing if \(A\) is included in the set \(\A\). Therefore, the total number of integers present in the array is \(\binom{7}{2} - |\A| = 28\).
	\begin{figure*}
			\begin{align}
			\label{eg algo 1 pda}
			\resizebox{\textwidth}{!}{$
			P_{7,2}=\begin{blockarray}{c|ccc|ccc|ccc|ccc|ccc|ccc|ccc}
				& &\{123\} && &\{145\}& & &\{167\}& & &\{246\}& & &\{257\}&& &\{347\}& & &\{356\}& \\ \hline
				& \{12\}&\{13\} &\{23\}& \{14\} & \{15\}& \{45\}& \{16\}& \{17\}& \{67\} & \{24\} & \{26\}& \{46\}& \{25\}& \{27\}& \{57\}& \{34\}& \{37\}& \{47\} & \{35\}& \{36\}& \{56\} \\ \hline
				\begin{block}{c|ccc|ccc|ccc|ccc|ccc|ccc|ccc}
				\{1\} & * & * & *  & *  & *  & * & * & * & * & 2 & 4 & 11 & 3 & 5 & 14 & 6 & 9 & 12 & 7 & 8 & 13\\
				\{2\} & * & * & *  & 2  & 3 & 20 & 4 & 5 & 25 & * & * & * & * & * & * & 16 & 19 & 22 & 17 & 18 & 23  \\
				\{3\} & * & * & *  & 6  & 7 & 26 & 8 & 9 & 31 & 16 & 18 & 27 & 17 & 19 & 30 & * & * & * & * & * & *\\
				\{4\} & 2 & 6 & 16 & *  & * & * & 11 & 12 & 34 & * & * & * & 20 & 22 & 33 & * & * & * & 26 & 27 & 32 \\
				\{5\} & 3 & 7 & 17 & *  & * & * & 13 & 14 & 35 & 20 & 23 & 32 & * & * & * & 26 & 30 & 33 & * & * & *  \\
				\{6\} & 4 & 8 & 18 & 11 & 13 & 32 & * & * & *  & * & * & * & 23 & 25 & 35 & 27 & 31 & 34 & * & * & *\\
				\{7\} & 5 & 9 & 19 & 12 & 14 & 33 & * & * & *  & 22 & 25 & 34 & * & * & * & * & * & * & 30 & 31 & 35\\
				\end{block}
			\end{blockarray}
			$}
			\end{align}
	\end{figure*}
Consider a system with \(\Lambda=7\) mapper nodes indexed by \([7]\) and \(K = 7\) reducer nodes indexed by \( A \in \A \).  
We partition \( N = 7 \) files into \( 7 \) batches \(\{B_{1},B_{2},B_{3},B_{4},B_{5},B_{6},B_{7}\}\), where each batch is defined as \( B_f =\{w_f\} \) for \( f \in [7] \).  

The row index \( f \in [7] \) represents the batch \( B_f \), and the set of all columns \(\{(A,U)\!:\! U \subseteq A, |U| = 2\}\) represents the reducer node \( A \).  
A \( * \) appears in the row indexed by \( B_f \) and the column indexed by \( (A,U) \) if and only if the reducer node \( A \) has access to the batch \( B_f \), for each \( f \in [7] \) and \( A \in \A \).
Each reducer node \( A \in \A \) is assigned \( Q/K = 1 \) output function.  
The indices of the output functions assigned to each reducer node \( A \) are given by:
$\W_{\{123\}} \!=\! \{1\},  \W_{\{145\}} \!=\! \{2\},  \W_{\{167\}} \!=\! \{3\},  \W_{\{246\}} \!=\! \{4\},
\W_{\{257\}} \!=\! \{5\},  \W_{\{347\}} \!=\! \{6\}, $ and $ \W_{\{357\}} \!=\! \{7\}.$
The batch assigned to each mapper node \( f \in [7] \) is given by \( M_f = \{B_f\} \), leading to a computation load of \( r=1 \).  
For each \( f \in [7] \), the mapper node \( f \) computes \( Q = 7 \) IVs for each assigned input file.  
Now, consider a MADC model where each reducer node \( A \) is connected to the mapper nodes in the set \( \{f: f \in A\} \).
It can be observed that the array \( P_{7,2} \) corresponds to this model since the set of all batches assigned to each reducer node \( A \) is given by:
$R_{A}  =  \cup_{f\in [7]:p_{f,(A,U)} = *} M_{f}=  \cup_{f\in [7]:p_{f,(A,U)} = *} B_{f}.$
Consider the first three columns of \( P_{7,2} \), indexed by \( (\{123\},\{12\}) \), \( (\{123\},\{13\}) \), and \( (\{123\},\{23\}) \). The set of all integers present in these columns is given by
$\s_{\{123\}} = \{2,3,4,5,6,7,8,9,16,17,18,19\},$
where $\s_A$ represents the set of all integers present in the columns associated with the reducer node $A$.
We concatenate the IVs for the output functions in \( \mathcal{W}_{\{123\}} \), which need to be computed by the reducer node \( \{123\} \) and can be retrieved from the files in \( B_{4} \), i.e., 
$\{v_{q,n} : q \in \W_{\{123\}}, w_n \in B_{4} \},$
into a single symbol:
$\U_{\W_{\{123\}},B_{4}} = (v_{q,n} : q \in \{1\}, w_n \in \{w_4\}).$
Next, we partition the symbol \( \U_{\W_{\{123\}},B_{4}} \) into \( 3 \) equal-sized packets:
$\U_{\W_{\{123\}},B_{4}}=\{\U_{\W_{\{123\}},B_{4}}^{\{12\}},\U_{\W_{\{123\}},B_{4}}^{\{13\}},\U_{\W_{\{123\}},B_{4}}^{\{23\}}\}.$
Similarly, we concatenate the IVs for the output functions in \( \W_{\{123\}} \), which need to be computed by the reducer node \( \{123\} \) and can be retrieved from the files in \( B_{5} \), i.e.,
$\{v_{q,n} : q \in \W_{\{123\}}, w_n \in B_{5} \},$
into the symbol:
$\U_{\W_{\{123\}},B_{5}} = (v_{q,n} : q \in \{1\}, w_n \in \{w_5\}).$
We then partition it into three equal-sized packets:
$\U_{\W_{\{123\}},B_{5}}=\{\U_{\W_{\{123\}},B_{5}}^{\{12\}},\U_{\W_{\{123\}},B_{5}}^{\{13\}},\U_{\W_{\{123\}},B_{5}}^{\{23\}}\}.$
This process is repeated for all other batches and reducer nodes.
Now, consider the entry \( s=2 \) in \( \s_{\{123\}} \). The other occurrences of the integer \( 2 \) appear in the columns indexed by \( (\{145\}, \{14\}) \) and \( (\{246\},\{24\}) \). Consequently, we further partition the symbols in \( \U_{\W_{\{123\}},B_{4}}^{\{12\}} \) into two equal-sized packets:
$\U_{\W_{\{123\}},B_{4}}^{\{12\}}=\{\U_{\W_{\{123\}},B_{4}}^{\{12\}, \{145\}},\U_{\W_{\{123\}},B_{4}}^{\{12\},\{246\}}\}.$
This partitioning process is repeated for all other integer entries in the array.
 Since $|\s_A|=12, \forall A\in \A$, each reducer node $A$ transmits 12 coded symbols $X_A^s$ for $s\in \s_{A}$.
	For each entry \( s \in \s_A \) (suppose $p_{\lambda, (A, U_A)} = s$), the reducer node \( A \in \A\) generates a coded symbol as follows:
	\begin{align}
	X_A^s = \bigoplus_{{\substack{(\lambda_i,A_i)\in [7] \times (\A \backslash A) :\\ p_{\lambda_i,(A_i,U_{A_i})} =s, \text{ for } U_{A_i} \subseteq A_i, |U_{A_i}| = 2}}} U_{\mathcal{W}_{A_i},B_{\lambda_i}}^{U_{A_i},A}.
	\end{align}
	The reducer node \( A \) then multicasts the sequence
	$
	\mathbf{X}_A = \{X_A^s : s \in \s_A\}.
	$
	The reducer node $\{123\}$ can retrieve $\U_{\W_{\{123\}},B_{4}}^{\{12\}}$  from the coded symbols $X_{\{145\}}^{2}$ and $X_{\{246\}}^{2}$  transmitted by the reducer nodes $\{145\}$ and $\{246\}$ respectively. Similarly, it can retrieve all the missing symbols to compute the function $\phi_1$. It can be verified that all other reducer nodes can retrieve all required symbols needed to compute the respective output functions. A total of $84$ coded symbols are transmitted across the reducer nodes, each of size $\beta/6$ bits.  Hence, the communication load is $L(1) =\frac{\frac{84\beta}{6}}{7*7*\beta}=\frac{2}{7}$.\qed
\end{example}
\begin{table}[!h]
	\setlength{\belowcaptionskip}{0pt} 
	\centering
	\small
	\begin{tabular}{ | c|c|c|}
		\hline	
		\textbf{Parameters}&  \textbf{Example \ref{exmp1} ($t$-design)} & \textbf{CT }\\  \hline \hline
		\textit{No. of mappers: $\Lambda$}  &7& 7\\ \hline
		\textit{ No. of reducers: $K$} &7&35 \\ \hline
		\textit{No. of batches: $F$}  &7&7 \\ \hline
		\textit{No. of files: $N$}  &7& 7\\ \hline
		\textit{No. of output functions: $Q$}  &7&35 \\ \hline
		\textit{Communication load: $L$} &0.28& 0.19\\ \hline
	\end{tabular}
	\caption{Comparison of $t$-design and CT for $\alpha = 3 $ and $r=1$.}
	\label{tab1}
\end{table}
Comparing Example \ref{exmp1} with NNC-MRG from \cite{SGR} for the parameters \(\Lambda =7\), \(r=1\), and \(\alpha=3\), we find that the scheme proposed in \cite{SGR} is not applicable. This is because the scheme relies on the condition that \((\Lambda - (\alpha -1)r)\) divides \(2\Lambda\), which does not hold in this case.  

Next, a comparison of Example \ref{exmp1} with the CT scheme is presented in Table \ref{tab1}. When examining Example \ref{exmp1} and CT under the same parameters (as listed in Table \ref{tab1}), we observe that both have identical values for the number of mapper nodes, computation load, \(\alpha\), and the total number of files. However, the CT scheme requires 35 reducer nodes and output functions, whereas Example \ref{exmp1} requires only 7. This highlights the advantage of using $t$-design in terms of reducing the number of reducer nodes, though it comes at the cost of  increased communication load.

\section{Proof of Theorem \ref{thm1}}
\label{proof thm1}
In this section, we demonstrate that the array generated by \textbf{Algorithm \ref{algo1}}  forms a \( g \)-regular MRA.

In \textbf{procedure 1}, we list all subsets of size \( t+1 \) from the set \( [\Lambda] \) in lexicographic order. For any subset \( T' \) of size \( t+1 \), we define \( y_{t+1}(T') \) as its position in the order. This function \( y_{t+1} \) is clearly a bijection from the set of all \( (t+1) \)-subsets of \( [\Lambda] \) to the index set \( \left[ {\Lambda \choose t+1} \right] \). For example, when \( \Lambda = 5 \) and \( t = 3 \), the subsets of size \( t+1 = 4 \) from \( [5] \) are arranged as follows:
$\{1234\}, \{1235\}, \{1245\}, \{1345\}, $ and $\{2345\}.$
Thus, the function \( y_4 \) assigns the values:
$y_4(\{1234\}) \!=\! 1,  y_4(\{1235\}) \!=\! 2,  y_4(\{1245\}) \!=\! 3, 
y_4(\{1345\}) \!=\! 4,$ and $ y_4(\{2345\}) \!=\! 5.$

In \textbf{procedure 2}, we define the array \( P_{\Lambda,t} \). The rows of this array are indexed by \( \{\lambda : \lambda \in [\Lambda]\} \), and the columns are indexed by \( \{(A, U) \!:\! A \in \mathcal{A}, U \subseteq A, |U| \!=\! t \} \). Hence, the number of rows in the array is $\Lambda$ and the number of columns is $|\A|{\alpha \choose t} =\frac{{\Lambda \choose t}}{{\alpha \choose t}}\times {\alpha \choose t} ={\Lambda \choose t}$.

Next, we need to verify that the array satisfies the conditions \( C1' \) and \( C2 \) from {\bf Definition \ref{def:g-SRA}}. From the structure of the array (as shown in eq. (\ref{Dk})), it is evident that the symbol \( * \) appears whenever \( |\lambda \cap A| \neq 0 \), i.e., when \( \lambda \) and \( A \) share some common element. Since the size of \( A \) is \( \alpha \), the symbol \( * \) will appear exactly \( \alpha \) times in each column indexed by \( (A, U) \).

Next, consider two distinct entries \( p_{\lambda_1, (A_1, U_1)} = p_{\lambda_2, (A_2, U_2)} = s \), where \( A_1, A_2 \in \mathcal{A} \), \( U_1 \subseteq A_1 \), \( U_2 \subseteq A_2 \) with \( |U_1| = |U_2| = t \), and \( \lambda_1, \lambda_2 \in [\Lambda] \). Given that \( s \) is an integer only if \( |\lambda_1 \cap A_1| = |\lambda_2 \cap A_2| = 0 \), and that \( y_{t+1} \) is a bijection from \( [\{T' \subseteq [\Lambda] : |T'| = t+1 \}] \) to \( \left[ {\Lambda \choose t+1} \right] \), we know that \( \lambda_1 \cup U_1 = \lambda_2 \cup U_2 \). By {\bf Definition \ref{def: t design}}, every \( t \)-subset of \( [\Lambda] \) in $t$-$(\Lambda,\alpha,1)$ design appears in exactly one block, which implies that
\begin{itemize}
	\item \( \lambda_1 \!\neq\! \lambda_2 \) and \( U_1 \!\neq\! U_2 \), meaning the two entries are in distinct rows and columns. 
	This condition is equivalent to \( |\lambda_1 \cap U_2| \!\neq\! 0 \) and \( |\lambda_2 \cap U_1| \!\neq\! 0 \) (since \( \lambda_1 \cup U_1 \!=\! \lambda_2 \cup U_2 \)). This  implies that \( |\lambda_1 \cap A_2| \!\neq\! 0 \) and \( |\lambda_2 \cap A_1| \!\neq\! 0 \). Thus, \( p_{\lambda_1, (A_2, U_2)} = p_{\lambda_2, (A_1, U_1)} = * \) by eq. (\ref{Dk}), satisfying condition \( C2 \) of {\bf Definition \ref{def:g-SRA}}.
\end{itemize}

Consider a \( (t+1)  \)-subset \( T' \) of \( [\Lambda] \) such that \( T' \subseteq A \) for some \( A \in \mathcal{A} \) (i.e., \( T' \) is contained in block \( A \)). For each \( t \)-subset \( U \) of \( T' \), we have \( p_{T' \backslash U, (A, U)} = * \), as given by eq. (\ref{Dk}). For instance, if \( A = \{1234\} \), \( t = 2 \), and \( T' = \{123\} \), then \( p_{1, (A, \{2 3\})} = p_{2, (A, \{13\})} = p_{3, (A, \{12\})} = * \).

Additionally, by {\bf Definition \ref{def: t design}}, these \( t \)-subsets \( U \) do not appear in any block other than \( A \), because in a $t$-$(\Lambda,\alpha,1)$ design, each \( t \)-subset \( U \) of \([ \Lambda ]\) is contained in exactly one block. This implies that the integer \( y_{t+1}(T') \) will not appear as an entry in the array \( P_{\Lambda, t} \). 
The total number of such integers \( S' \), which do not appear in \( P_{\Lambda, t} \), corresponds to the total number of \( (t+1) \)-subsets \( T' \) of \( [\Lambda] \) such that \( T' \subseteq A \) for some \( A \in \mathcal{A} \). Thus, we compute
	$S' = |\A| {\alpha \choose t+1} = \frac{{\Lambda \choose t} {\alpha \choose t+1}}{{\alpha \choose t}} = {\Lambda \choose t} \left( \frac{\alpha-t}{t+1} \right).$

Next, we consider the \( (t+1) \)-subsets \( T' \) of \( [\Lambda] \) such that \( T' \nsubseteq A \) for any \( A \in \mathcal{A} \). For each \( t \)-subset \( U \) of \( T' \), the definition of the \( t \)-design guarantees the existence of exactly one block \( A \) such that \( U \in A \), and \( p_{T' \backslash U, (A, U)} = y_{t+1}(T') \). 

The mapping \( y_{t+1} \) assigns a unique integer to each \( (t+1)  \)-subset \( T' \). The structure of the array ensures that \( y_{t+1}(T') \) appears only in rows and columns corresponding to \( t \)-subsets \( U \) that satisfy the \( t \)-design condition. Since \( U \) is contained in exactly one block \( A \), the integer \( y_{t+1}(T') \) appears exactly \( t+1 \) times in the array, once for each \( t \)-subset \( U \) of \( T' \). 
The total number of integers present in the array is 
	$S = {\Lambda \choose t+1} - {\Lambda \choose t} \left( \frac{\alpha-t}{t+1} \right) 
	= {\Lambda \choose t} \left( \frac{\Lambda-t}{t+1} - \frac{\alpha-t}{t+1} \right) 
	= {\Lambda \choose t} \left( \frac{\Lambda-\alpha}{t+1} \right).$
Since each integer appears \( t+1 \) times in the array, condition \( C1' \) is satisfied.
In summary, both conditions \( C1' \) and \( C2 \) are satisfied, completing the proof.

\section{Proof of Theorem \ref{thm2}}
\label{proof thm2}

 In this section, we describe how a MADC scheme can be constructed for an MRG with $\Lambda$ mapper nodes and $K = |\mathcal{A}|$ reducer nodes, where each reducer node is connected to a subset of $\alpha$ mapper nodes. The construction is based on a $t$-$(\Lambda, \alpha, 1)$ design, denoted as $([\Lambda], \mathcal{A})$.
We consider $Q = \eta_2 K$ output functions, where $\eta_2$ is an integer and each reducer node is assigned $\eta_2$ output functions for computation.
\subsection{Map Phase}
	The $N$ files are divided into $\Lambda$ disjoint batches $\{B_1, B_2, \ldots, B_{\Lambda}\}$, with each batch containing $\eta_1 = N / \Lambda$ files. Thus, we have
	$
	\bigcup_{m=1}^{\Lambda} B_m = \{w_1,  \ldots, w_N\}.
	$
	We construct  $(t+1)$-$ \left (  {\Lambda \choose t},\Lambda,  {\Lambda \choose t} (  \frac{\Lambda-\alpha}{t+1} ) \right )$  MRA $P_{\Lambda, t}$ using {\bf Algorithm \ref{algo1}}.
	The MRA has $|\A|{\alpha \choose t} ={\Lambda \choose t}$ columns, indexed by $(A, U)$, where $A \in \mathcal{A}$ and $U \subseteq A$ with $|U| = t$.  
	\begin{itemize}
		\item For each $A \in \A$, the set of ${\alpha \choose t}$ columns indexed as $\{(A, U) : U \subseteq A, |U| = t\}$ corresponds to the reducer node $A$.
		\item Each mapper node $\lambda \in [\Lambda]$ is assigned a single batch, $M_\lambda = \{B_\lambda\}$, making the computation load $r = 1$.
		\item 	Each batch $B_\lambda$ corresponds to a row in the MRA.
	\end{itemize}
	For each $\lambda \in [\Lambda]$, mapper node $\lambda$ computes the IVs
	$\{v_{q,n} : q \in [Q], w_n \in B_\lambda\}$
	where each $v_{q,n}$ is a bitstream of length $\beta$.
	Each reducer node $A \in \mathcal{A}$ is connected to mapper node $\lambda$ if $\lambda \in A$. Hence, each reducer node $A$ can access all batches in the set
$R_A = \{B_\lambda : \lambda \in A\}.$

\subsection{Shuffle Phase}
	Each reducer node $A \in \mathcal{A}$ retrieves the IVs:
		$ \{v_{q,n} : q \in [Q], w_n \in B_{\lambda},  \lambda \in A\}.$
	If the entry in the MRA corresponding to row $\lambda$ and column $(A, U)$ is
		 $p_{\lambda, (A, U)} = *$, it indicates that reducer $A$ has access to batch $B_\lambda$, for  $\lambda \in [\Lambda],A\in \A$ and $U \subseteq A$ with $|U| =t.$

Consider the row indexed by \(\lambda\) and the columns indexed by \(\{(A, U) : U \subseteq A, |U| = t\}\), i.e., the columns representing the reducer node \(A\) such that \(\lambda \cap A \neq 0\).  
We concatenate the set of IVs corresponding to the output functions in \(\mathcal{W}_A\), which need to be computed by the reducer node \(A\) and can be obtained from the files in \(B_{\lambda}\). This concatenated set is represented as 
\begin{align}
\label{symbols_pda}
\U_{\mathcal{W}_A,B_{\lambda}} = (v_{q,n} : q \in \mathcal{W}_{\lambda}, w_n \in B_{\lambda}) \in \F_{2^{\eta_1 \eta_2 \beta}}.
\end{align}
Next, we partition the symbols \(\U_{\mathcal{W}_A, B_{\lambda}}\) into \(\binom{\alpha}{t}\) equal-sized packets such that
\begin{align}
\label{partition first}
\U_{\mathcal{W}_A,B_{\lambda}}=\{\U_{\mathcal{W}_A,B_{\lambda}}^{U_A} : U_A \subseteq A, |U_A|=t\}.
\end{align}
Consider an entry in the MRA \( P_{\Lambda,t} \) corresponding to the row indexed by \(\lambda\) and the column indexed by \((A, U_A)\), given by \(p_{\lambda, (A, U_A)} = s\), where \(s\) is some integer. Each integer \(s\) appears exactly \(t+1\) times in the MRA. Let the other \(t\) occurrences of \(s\) be represented as follows  
\begin{align}
p_{\lambda_1, (A_1, U_{A_1})} = p_{\lambda_2, (A_2, U_{A_2})} = \cdots = p_{\lambda_t, (A_t, U_{A_t})} = s.
\end{align}
For each \(i \in [t]\), it is known from Section \ref{proof thm1} that \(p_{\lambda, (A_i, U_{A_i})} = *\).
We then further partition the symbol \(\U_{\mathcal{W}_A, B_{\lambda}}^{U_A}\) into \(t\) equal-sized packets such that
\begin{align}
\label{partition symbols}
\U_{\mathcal{W}_A,B_{\lambda}}^{U_A}=\{\U_{\mathcal{W}_A,B_{\lambda}}^{U_A,A_1}, \U_{\mathcal{W}_A,B_{\lambda}}^{U_A,A_2}, \ldots, \U_{\mathcal{W}_A,B_{\lambda}}^{U_A,A_{t}}\}.
\end{align}
Let \( \s_A \) denote the set of distinct integers appearing in the columns associated with \( A \in \mathcal{A} \), i.e., the columns corresponding to \( \{(A, U) : U \subseteq A, |U| = t\} \). For each entry \( s \in \s_A \) (suppose $p_{\lambda, (A, U_A)} = s$), the reducer node \( A \) generates a coded symbol as follows
\begin{align}
	\label{transmission}
	X_A^s = \bigoplus_{{\substack{(\lambda_i,A_i)\in [\Lambda] \times (\A \backslash A) :\\ p_{\lambda_i,(A_i,U_{A_i})} =s, \text{ for } U_{A_i} \subseteq A_i, |U_{A_i}| = t}}} U_{\mathcal{W}_{A_i},B_{\lambda_i}}^{U_{A_i},A}.
\end{align}
The reducer node \( A \) then multicasts the sequence $\mathbf{X}_A = \{X_A^s : s \in \s_A\}.$ 
The reducer node \( A \) can compute the coded symbol \( X_A^s \) using the IVs that are accessible to it. Specifically, for each \((\lambda_i, A_i)\) in the summation above, \( p_{\lambda_i, (A_i, U_{A_i})} = p_{\lambda, (A, U_A)} = s \). Since \( A \neq A_i \), we also have \( \lambda \neq \lambda_i \), and from Section \ref{proof thm1}, \( p_{\lambda_i, (A, U_A)} = * \).
Thus, the reducer node \( A \) has access to the IVs \( \{v_{q, n} : q \in \mathcal{W}_{A_i}, w_n \in B_{\lambda_i}\} \). Consequently, it can construct the symbol \( U_{\mathcal{W}_{A_i}, B_{\lambda_i}}^{U_{A_i},A} \). 


\subsection{Reduce Phase}
Upon receiving the sequences \( \{ \mathbf{X}_{A'} \}_{A' \in \mathcal{A} \setminus A} \), each reducer node \( A \in \mathcal{A} \) decodes all IVs associated with its output functions, i.e., \( \{ v_{q,n} : q \in \mathcal{W}_A, n \in [N] \} \), using the IVs 
 that are accessible to it. 
Specifically, the reducer node \( A \) needs to compute the set of IVs \( \{ v_{q,n} : q \in \W_A, w_n \in B_{\lambda}, B_{\lambda} \notin R_A \} \), which are the IVs required for the output functions in \( \mathcal{W}_A \) from files that it does not have access to (i.e., from the files in \( B_{\lambda} \) where \( \lambda \in [ \Lambda] \) and \( p_{\lambda, (A, U_A)} \neq * \)).

Without loss of generality, assume that \( p_{\lambda, (A, U_A)} = s \in \s_A \). For each \( i \in [t] \), the reducer node \( A \) can compute the symbol \( U_{\mathcal{W}_A, B_{\lambda}}^{U_{A},A_i} \) in (\ref{partition symbols}) from the coded symbol \( X_{A_i}^s \) transmitted by the reducer node \( A_i \), i.e.,
\begin{align}
	\label{node l transmission}
	X_{A_i}^s = \bigoplus_{{\substack{(e, v) \in [\Lambda] \times (\A \setminus A_i) :\\ p_{e,(v,u)} =s, \text{ for } u \subseteq v, |u| = t}}} U_{\mathcal{W}_v, B_e}^{u,A_i}.
\end{align}
In (\ref{node l transmission}), for \( v \neq A \), \( p_{e, (v, u)} =p_{\lambda, (A, U_A)} = s \) implies that \( p_{e, (A, U_A)} = * \) from Section \ref{proof thm1}. Thus, the reducer node \( A \) can compute \( U_{\mathcal{W}_v, B_e}^{u,A_i} \) using the relations in  (\ref{symbols_pda}), (\ref{partition first}) and (\ref{partition symbols}). For \( v = A \), \( p_{e, (v, u)} =p_{\lambda, (A, U_A)} = s \) implies \( e = \lambda \) and $u=U_A$ (from Section \ref{proof thm1}). Therefore, the reducer node \( A \) can recover the symbol \( U_{\mathcal{W}_A, B_{\lambda}}^{U_{A},A_i} \) from the coded symbol in (\ref{node l transmission}) by canceling out the other terms. By collecting all symbols \( U_{\mathcal{W}_A, B_{\lambda}}^{U_{A},A_i} \) from (\ref{partition symbols}), the reducer node \( A \) can successfully compute the output functions in \( \mathcal{W}_A \). 

Next, we compute the communication load for this scheme. For each integer \( s \) in the array \( P_{\Lambda, t} \), there are \( t+1 \) associated sequences sent, each of size \( \frac{\eta_1 \eta_2 \beta}{{\alpha \choose t}t} \) bits, as per (\ref{transmission}). Since there are \( S={\Lambda \choose t} \left( \frac{\Lambda-\alpha}{t+1} \right)\) distinct integers in the array, each appearing exactly \( t+1 \) times, the total communication load is given by:
		\begin{align}
	L &= \frac{1}{QN\beta} \frac{ (t+1)S \eta_1 \eta_2 \beta}{{\alpha \choose t}(t)} 
	= \frac{{\alpha \choose t}}{\Lambda {\Lambda \choose t}} \frac{{\Lambda \choose t}(\Lambda-\alpha)}{{\alpha \choose t}t}  
	=  \left ( \frac{\Lambda-\alpha}{\Lambda t}\right ).
\end{align}


\begin{thebibliography}{9}
	\bibitem{Mapreduce}
	J. Dean and S. Ghemawat, ``Mapreduce: Simplified data processing on large clusters," in \emph{Commun. ACM}, vol. 51, no. 1, pp. 107-113, Jan. 2008.
	\bibitem{LMA}
	S. Li, M. A. Maddah-Ali, Q. Yu and A. S. Avestimehr, ``A Fundamental Tradeoff Between Computation and Communication in Distributed Computing," in \emph{IEEE Trans. Inf. Theory}, vol. 64, no. 1, pp. 109-128, Jan. 2018.
	\bibitem{YYW}
	Q. Yan, S. Yang and M. Wigger, ``Storage computation and communication: A fundamental tradeoff in distributed computing", in \emph{Proc. IEEE Inf. Theory Workshop}, pp. 1-5, Guangzhou, China, Sep. 2018.
	\bibitem{SFZ}
	L. Song, C. Fragouli and T. Zhao, ``A pliable index coding approach to data shuffling", in \emph{IEEE Trans. Inf. Theory}, vol. 66, no. 3, pp. 1333-1353, Mar. 2020.
	\bibitem{YL}
	H. Yang and J. Lee, ``Secure distributed computing with straggling servers using polynomial codes", in \emph{IEEE Trans. Inf. Forensics Security}, vol. 14, no. 1, pp. 141-150, Jan. 2019.
	\bibitem{DPYTH}
	S. Dhakal, S. Prakash, Y. Yona, S. Talwar and N. Himayat, ``Coded computing for distributed machine learning in wireless edge network", in \emph{Proc. IEEE Veh. Technol. Conf.}, Honolulu, HI, Sep. 2019, pp. 1-6.
	\bibitem{LCW}
	F. Li, J. Chen and Z. Wang, ``Wireless MapReduce distributed computing", in \emph{IEEE Trans. Inf. Theory}, vol. 65, no. 10, pp. 6101-6114, Oct. 2019.
	\bibitem{WCJ}
	N. Woolsey, R.-R. Chen and M. Ji, ``A combinatorial design for cascaded coded distributed computing on general networks", in \emph{IEEE Trans. Commun.}, vol. 69, no. 9, pp. 5686-5700, Sep. 2021.
	\bibitem{LMAFog}
	S. Li, M. A. Maddah-Ali and A. S. Avestimehr, ``Coding for distributed fog computing", in \emph{IEEE Commun. Mag.}, vol. 55, no. 4, pp. 34-40, Apr. 2017.
	\bibitem{PLSSM}
	H. Park, K. Lee, J. Sohn, C. Suh and J. Moon, ``Hierarchical coding for distributed computing", in \emph{Proc. IEEE Int. Symp. Inf. Theory}, Vail, CO, June 2018, pp. 1630-1634.
	\bibitem{WCJnew}
	N. Woolsey, R. Chen and M. Ji, ``A new combinatorial design of coded distributed computing", in \emph{Proc. IEEE Int. Symp. Inf. Theory}, Vail, CO, June 2018, pp. 726-730.
	\bibitem{JQ}
	J. Jiang and L. Qu, ``Cascaded coded distributed computing schemes based on placement delivery arrays", in \emph{IEEE Access,} vol. 8, pp. 221385-221395, Dec. 2020.
	\bibitem{YTC}
	Q. Yan, X. Tang and Q. Chen, ``Placement delivery array and its applications", in \emph{Proc. IEEE Inf. Theory Workshop}, pp. 1-5, Guangzhou, China, Nov. 2018.
		\bibitem{SG}
	S. Sasi and O. Günlü, ``Rate-Limited Shuffling for Distributed Computing,'' \textit{Proc. IEEE Int. Sym. Inf. Theory}, Athens, Greece, pp. 2778-2783, July 2024.
	\bibitem{SG2}
	------, ``Secure Coded Distributed Computing,'' \textit{Proc. IEEE Int. Workshop Sig. Process. Adv. Wireless Commun.}, Lucca, Italy, pp. 811-815, Sep. 2024.
	\bibitem{SG3}
	------, ``Secure Coded Distributed Computing and Extensions to Multiple Access Setting,'' \textit{IEEE Trans. Commun. (TCOM)}, doi: 10.1109/TCOMM.2025.3569733.
	\bibitem{SG4}
	------, ``Private Coded Distributed Computing Framework,'' in \textit{Proc. IEEE Int. Sym. Inf. Theory (ISIT)}, Ann Arbor, Michigan, USA, June
	2025.
	\bibitem{YCTCPDA} 
	Q. Yan, M. Cheng, X. Tang and Q. Chen, ``On the Placement Delivery Array Design for Centralized Coded Caching Scheme,'' in \emph{IEEE Trans. Inf. Theory}, vol. 63, no. 9, pp. 5821-5833, Sep. 2017.
	\bibitem{BP}
	B. Federico and P. Elia, ``Multi-Access Distributed Computing,'' in \emph{IEEE Trans. Inf. Theory}, vol. 70, no. 5, pp. 3385-3398, May 2024.
	\bibitem{SGR}
	S. Sasi, O. Günlü and B. S. Rajan, ``Multi-access Distributed Computing Models using Map-Reduce Arrays,'' \textit{Proc. IEEE Int. Sym. Inf. Theory}, Athens, Greece, pp. 1355-1360, July 2024.
		
	\bibitem{stinson}
	D. R. Stinson, \textit{Combinatorial designs: constructions and analysis}. Springer, NY, 2004.
	\bibitem{design1}
	J. Li and Y. Chang, ``Placement Delivery Arrays Based on Combinatorial Designs," in \textit{IEEE Commun. Letters}, vol. 26, no. 2, pp. 296-300, Feb. 2022.
	\bibitem{design2}
	M. Cheng, K. Wan, P. Elia and G. Caire, ``Coded Caching Schemes for Multi-Access Topologies via Combinatorial Design Theory,'' \textit{Proc. IEEE Int. Sym. Inf. Theory}, Taipei, Taiwan, Jun. 2023, pp. 144-149.

\end{thebibliography}
\end{document}